\documentclass[traditabstract]{aa}  

\usepackage{psfig}

\begin{document}

\title{Is CGCS 5926 a symbiotic X--ray binary?\thanks{Partly 
based on observations collected at the Astronomical Observatories of 
Asiago and Loiano (Italy).}}


\author{N. Masetti\inst{1}, 
U. Munari\inst{2},
A.A. Henden\inst{3},
K.L. Page\inst{4},
J.P. Osborne\inst{4} and
S. Starrfield\inst{5}
}

\institute{
INAF -- Istituto di Astrofisica Spaziale e Fisica Cosmica di 
Bologna, Via Gobetti 101, I-40129 Bologna, Italy
\and
INAF -- Osservatorio Astronomico di Padova, Sede di Asiago,
Via dell'Osservatorio 8, I-36012 Asiago, Italy
\and
AAVSO, 49 Bay State Road, Cambridge, MA 02138, USA
\and
Department of Physics \& Astronomy, University of Leicester, 
Leicester, LE1 7RH, United Kingdom
\and
School of Earth and Space Exploration, Arizona State University, 
Tempe, AZ 85287-1404, USA
}

\offprints{N. Masetti (\texttt{masetti@iasfbo.inaf.it)}}
\date{Received 15 May 2011; accepted 12 September 2011}

\abstract{We here report on multiwavelength (X--ray to optical) followup 
observations of carbon star CGCS 5926. These were motivated by the fact 
that this star is positionally coincident with a faint X--ray emitting 
object belonging to the {\it ROSAT} catalog of sources, thus suggesting a 
possible symbiotic X--ray binary (SyXB) nature for it. Our 
spectrophotometric optical data confirm the giant carbon star nature of 
the object and allow us to classify its spectral type as C(6,2). This 
classification places CGCS 5926 at a distance of $\sim$5 kpc from Earth. 
$BVR_{\rm C}I_{\rm C}$ photometry of the star shows that it displays a 
variability of $\sim$0.3 mag on timescales of months, with the star 
getting bluer when its brightness increases. Our photometric data indicate 
a periodicity of 151 days, which we explain as due to radial pulsations of
CGCS 5926 on the basis of its global characteristics. 
The source is not detected at X--rays with {\it Swift}/XRT down to 
a 0.3--10 keV band luminosity of $\approx$3$\times$10$^{32}$ erg s$^{-1}$. 
This nondetection is apparently in contrast with the {\it ROSAT} data; however
we show that, even if the probability that CGCS 5926 can be a SyXB appears
quite low, the present information does not completely rule out 
such a possibility, while it makes other interpretations even more 
unlikely if we assume that the {\it ROSAT} detection was real. 
This issue might thus be settled by future, more sensitive, observations 
at high energies.}


\titlerunning{is star CGCS 5926 a SyXB?}
\authorrunning{N. Masetti et al.}

\maketitle

\section{Introduction}

Low mass X--ray binaries (LMXBs) are interacting systems composed of
a compact object, neutron star (NS) or black hole, which is accreting 
from a late-type companion star, with mass generally $\la$1 $M_\odot$
and still on the main sequence (or possibly slightly evolved).

There exists however a handful of cases in which the donor star is actually
a red giant: these, by analogy with the symbiotic binaries which are 
formed by an evolved late-type star and a white dwarf, are dubbed 
symbiotic X--ray binaries (SyXBs; see e.g. Masetti et al. 2006a).

Observationally, these systems are characterized by appreciable X--ray 
emission ($\sim$10$^{32}$--10$^{34}$ erg s$^{-1}$; see Masetti et al. 2007 
and references therein) positionally associated with a red giant star 
which spectroscopically does not show any abnormal features, with the 
possible exception of a continuum excess in the blue and ultraviolet 
ranges. A notable outlier is the source GX 1+4, which emits up to 
$\sim$10$^{37}$ erg s$^{-1}$ and shows several emission lines in the 
optical spectrum of the red giant companion (Chakrabarty \& Roche 1997;
Munari \& Zwitter 2002). 
This is most likely due to the fact that, in this latter case, accretion 
onto the compact object takes place via Roche lobe overflow rather than 
via a stellar wind (but see Hinkle et al. 2006 for a different scenario). 
This allows the creation of a disk around the accretor and makes the mass 
transfer phenomenon more efficient in terms of production and reprocessing 
of X--ray emission.

Moreover, X--ray pulsations, with periods ranging from hundreds up to tens 
of thousands of seconds, were detected from these systems: this indicates
that the accreting compact object is a slowly rotating neutron 
star, the most extreme case being 4U 1954+319 ($P_{\rm spin} \sim$ 
18400 s; Corbet et al. 2008).

All the above characteristics make these systems rather unusual and, 
indeed, they are quite rare when compared to the number of known LMXBs 
(about 190, according to Liu et al. 2007): up to now, only 6 
systems are firmly included in this subclass of LMXBs (see Masetti et al. 
2007, Nespoli et al. 2010 and references therein) through coincidence 
between optical and X--ray positions and subsequent confirmation via optical 
or near-infrared (NIR) spectroscopy.
Three more cases however exist, having either the confirmation 
still pending due to the lack of optical or NIR spectroscopy (1RXS 
J180431.1$-$273932: Nucita et al. 2007; 2XMM J174016.0$-$290337: 
Farrell et al. 2010), or with a still debated nature (IGR J16393$-$4643: 
see Nespoli et al. 2010, but also Thompson et al. 2006 and Corbet et al. 
2010 for a different source classification as a supergiant high-mass 
X--ray binary).

Therefore, given the small number of known SyXBs, each new possible 
member of this subclass of LMXBs should be the object of an in-depth
multiwavelength study to expand the sample.

We thus focused our attention on star CGCS 5926, which was classified 
by Maehara \& Soyano (1987) as a $V$ = 14.8 mag carbon star (thus a 
late-type giant) in the Cassiopeia region; according to the 2MASS 
catalog (Skrutskie et al. 2006), this object has NIR magnitudes $J$ = 
8.824$\pm$0.023, $H$ = 7.490$\pm$0.042 and $K_{\rm s}$ = 6.942$\pm$0.024. 
Our interest in this star was drawn by the fact that (see Fig. 1) it is 
positionally inside the 20$''$ error circle of the soft X--ray source 1RXS 
J234545.9+625256 belonging to the {\it ROSAT} All-Sky Survey Faint Source 
Catalog (Voges et al. 2000), which makes CGCS 5926 a SyXB candidate due to 
its optical spectral classification and its possible X--ray emission.

However, the lack of further information at optical, X--ray and other
wavelengths does not make a secure case for inclusion of this star 
in the SyXB subclass. It is therefore worthy of further analysis 
by means of a specific spectrophotometric campaign in the optical range, 
as well as of a pointed observation with the use of an X--ray satellite 
affording localizations with a precision better than a few arcseconds.

We here report the results of optical, ultraviolet and X--ray observations 
of CGCS 5926 and a discussion of them, and we conclude that this star is 
either an X--ray variable or eclipsing SyXB, or that the low-significance 
{\it ROSAT} detection was spurious.

The outline of the paper is as follows: in Sect. 2 we describe our optical
and X--ray observations, while Sect. 3 reports the results and Sect. 4
a discussion of them. Finally, in Sect. 5 we present the conclusions of
our multiwavelength investigation of star CGCS 5926.

\begin{figure}[!t]
\psfig{figure=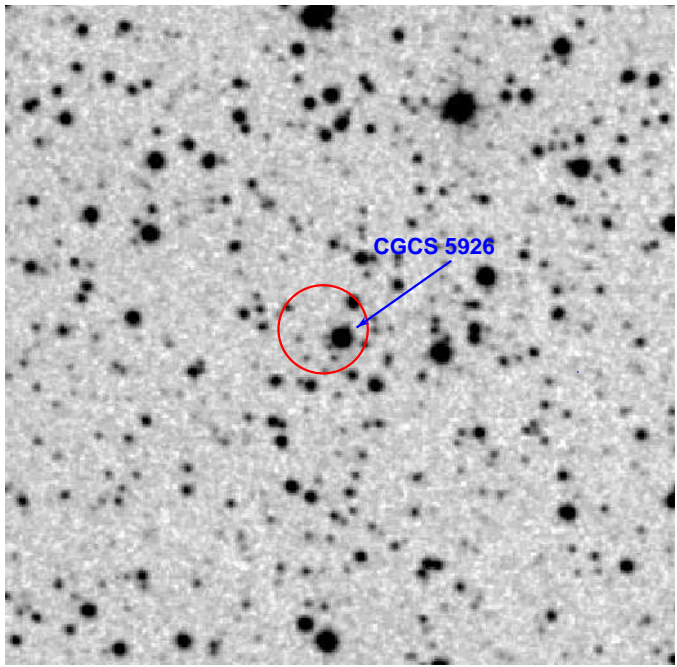,width=8.9cm,angle=0}
\caption[]{DSS-II-Red image of the field of CGCS 5926 (the star is indicated 
by the arrow) with superimposed the 20$''$-radius 0.1--2.4 keV {\it ROSAT} 
X--ray error circle. In the figure, North is at top and East is to 
the left. The field size is 5$'$$\times$5$'$.}
\end{figure}

\section{Observational data}

\subsection{Optical}

Optical spectroscopy of CGCS 5926 was acquired on 14 October 2009 with the 
1.82-metre ``Copernicus'' telescope of the Asiago Astronomical Observatory 
(Italy) plus the AFOSC instrument (which carries a 1024$\times$1024 pixel 
Tektronix TK1024 CCD). Using the Grism \#4 and a 1$\farcs$26 slit, this 
setup provided a dispersion of 4.2 \AA/pixel and a nominal 
coverage between 3500 and 7800 \AA. The total exposure time was 
2$\times$20 min centered at 02:03 UT.
Spectroscopic observations to further explore the blue part of the 
optical range were secured at the Astronomical Observatory of Bologna in 
Loiano (Italy) with the 1.52-metre ``Cassini" telescope plus the BFOSC 
instrument carrying an EEV 1300$\times$1340 pixel CCD: with the use of 
Grism \#6 and a slit width of 2$''$ we obtained a dispersion of 1.8 
\AA/pixel in the nominal range 3000--5300 \AA. Precisely, two 30-min 
spectra were obtained at mid-exposure time 02:13 UT of 9 August 2011.

Few additional 5500--8500 \AA~spectra with dispersion 1.1 \AA/pix were 
moreover obtained on 23 and 28 June 2011 and on 1 and 7 July 2011 with the
0.6-metre and 0.7-metre reflector telescopes operated by the ANS 
Collaboration\footnote{{\tt http://www.pd.astro.it/simbio-asiago/} (in 
italian)} in La Polse di Cougnes (Udine, Italy) and at the Schiaparelli 
Observatory in Campo de' Fiori (Varese, Italy), respectively, to check 
for possible presence of H$_\alpha$ emission, and confirming on all dates 
its absence.

The spectra, after correction for flat-field, bias and cosmic-ray
rejection, were background subtracted and optimally extracted (Horne 1986)
using IRAF\footnote{IRAF is the Image Analysis and Reduction Facility made
available to the astronomical community by the National Optical Astronomy
Observatories, which are operated by AURA, Inc., under contract with the
U.S. National Science Foundation. It is available at {\tt
http://iraf.noao.edu/}}. Wavelength calibration was performed using
comparison lamps acquired soon after each on-target spectroscopic
exposure, while flux calibration was accomplished by observing stars HR 
8780 and HR 8634 from the Asiago internal list of bright
spectrophotometric standard stars (Munari 2012, in preparation).
Wavelength calibration uncertainty was $\sim$0.5 \AA; this was checked by 
using the positions of background night sky lines. The flux calibration
of the Asiago spectra was verified against the source photometry 
acquired for this source (see below) and we found that it is correct to 
better than 5\% across the optical continuum; the flux scale of
the other spectra agrees with that of the Asiago ones. Spectra from a same 
observatory were then stacked together to increase the final 
signal-to-noise ratio.

Imaging data of the source were obtained with the robotic 14-inch 
Celestron telescope of the Sonoita Research Observatory (New Mexico, USA) 
in 2009 (from 15 October to 27 December) and 2010 (from 5 September 
to 17 November). The entire log of these observations is presented in 
Table 1. $BVR_{\rm C}I_{\rm C}$ Optec photometric filters were used; the 
detector was a 1024$\times$1024 pixel SBIG STL-1001E CCD camera, with a 
20$'$$\times$20$'$ field of view and a plate scale of 1$\farcs$25 
pix$^{-1}$. The frames were corrected for bias and flat-field and reduced 
by means of simple aperture photometry. Photometric calibration was 
obtained through continous monitoring of several equatorial standards 
(Landolt 1992) during the nights of observations.

\subsection{X--rays}

The field of CGCS 5926 was observed on 6 January 2010 in the 0.3--10 keV 
band with the X--Ray Telescope (XRT; Burrows et al. 2005) on board the 
{\it Swift} satellite (Gehrels et al. 2005). The XRT pointing (ID: 
00031572001) started at 08:34 UT, and 4374 s of on-source data were 
collected. The XRT data reduction was performed using the {\sc xrtdas} 
standard data pipeline package ({\sc xrtpipeline} v. 0.12.6), in order to 
produce screened event files. All data were extracted only in the photon 
counting (PC) mode (Hill et al. 2004), adopting the standard grade 
filtering (0--12 for PC) according to the XRT nomenclature, and using an 
extraction radius of 24$''$.

Within the observation we analysed, with the tool {\sc ximage} v. 4.4.1, 
the 0.3--10 keV image to search for sources detected (at a confidence 
level $>$3$\sigma$) both at the optical position of star CGCS 5926 and 
inside the {\it ROSAT} error circle.

\begin{figure*}[th!]
\vspace{-5cm}
\hspace{-1.5cm}
\psfig{figure=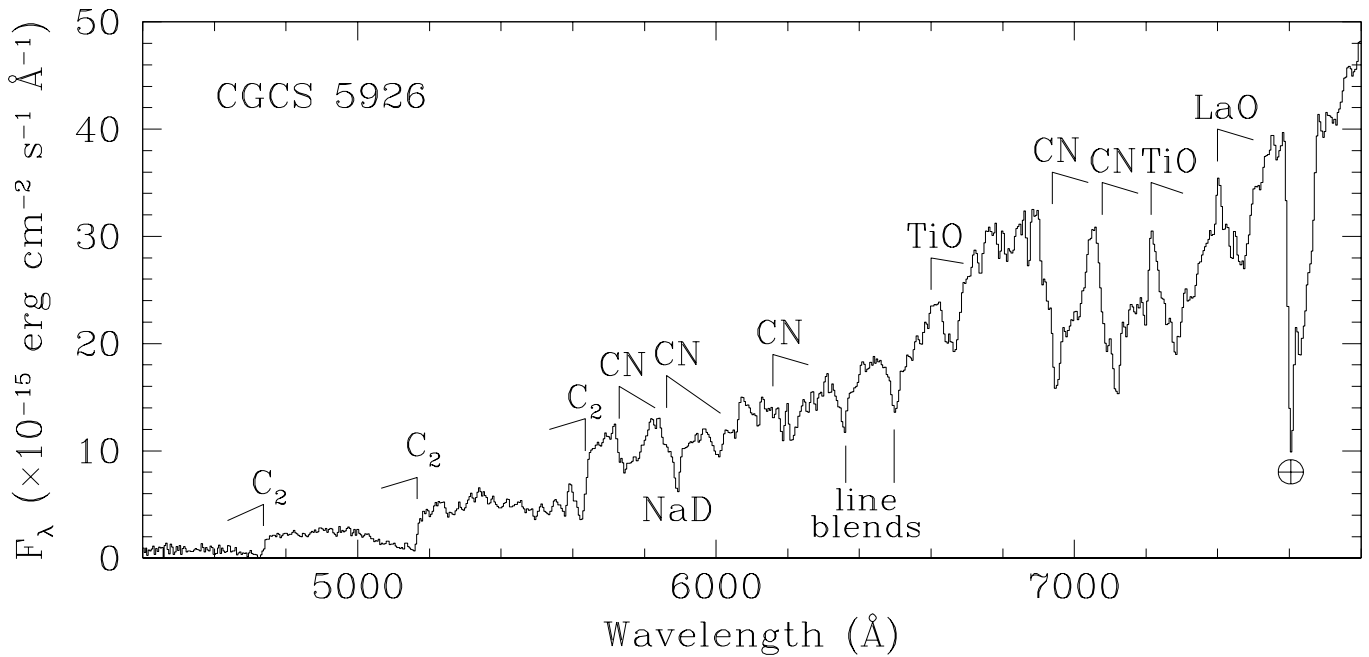,width=21cm,angle=-90}
\vspace{-1.3cm}
\caption[]{4400--7800 \AA~optical spectrum of star CGCS 5926 obtained with 
the Copernicus telescope on 14 October 2009. The spectrum is typical of a 
late-type giant carbon star (see text). The telluric absorption band at 
7600 \AA~is marked with the symbol $\oplus$.}
\end{figure*}

\subsection{Ultraviolet}

In parallel with the X--ray pointing, the UltraViolet-Optical Telescope 
(UVOT; Roming et al. 2005) onboard {\it Swift} as well, observed CGCS 5926 
in the $UVM2$ band ($\lambda$ = 2246 \AA; full width at half maximum: 498 
\AA) for a total of 4364 s starting at 08:37:42 UT of 6 January 2010. 
Count rates were measured through aperture photometry using 5$''$ 
apertures and were calibrated using the UVOT photometric system described 
by Poole et al. (2008).

\section{Results}

The optical spectrum of CGCS 5926 (Figs. 2 and 3) clearly shows the 
typical features of a carbon star (Yamashita 1967; Cohen et al. 1996): it 
is unmistakably dominated by the C$_2$ Swan bands at 4737, 5165 
and 5635 \AA~in the blue, and by CN bands redwards. We also find, among 
the main spectral features, the Na doublet at 5890 \AA~and two atomic line 
blends of metal intersystem lines of Fe {\sc i}, Ti {\sc i}, Cr {\sc i}, 
Ba {\sc i}, Ca {\sc i}, Mn {\sc i}, Co {\sc i} and Ni {\sc i} located at 
6352 \AA~and 6497 \AA~(see e.g. Turnshek et al. 1985). A telluric 
absorption feature is present at 7605 \AA. No emission features typical of 
X--ray binaries, such as Balmer and He {\sc ii} lines are present. In 
particular, no H$_\alpha$ or H$_\beta$ lines are readily detected either 
in emission or in absorption.

Moreover, the spectrum shown in Fig. 3 and covering the blue range of CGCS 
5926 clearly indicates that there is no evident excess in the $U$ band, as 
no signal is detected from this object blueward of $\sim$4200 \AA. This 
confirms the lack of signal detection in the blue part of the Asiago 
spectra acquired earlier, which is however not surprising given the 
red giant nature of the star coupled with the large optical absorption we 
infer toward it (see below).

Using the two-dimensional C($m$,$n$) diagnostics of Yamashita (1967) and 
the quantitative classification scheme derived by Cohen (1979), we can 
better classify the spectral type of CGCS 5926. Concerning the $m$ 
parameter, which is associated with the temperature index $T$, we find 
from our optical spectrum of CGCS 5926 shown in Fig. 2 that $T$ = 
0.45$\pm$0.03: this, from Table 2 of Cohen (1979) implies that $m$ = 6,
which corresponds to an effective blackbody temperature $T_{\rm eff} 
\approx$ 2500 K (Cohen 1979; Bergeat et al. 2001).
Likewise, the carbon indices $C1$, $C2$ and $C3$ are 0.51$\pm$0.04, 
0.58$\pm$0.04 and 0.90$\pm$0.06, respectively, which allow us to infer
that $n$ = 2 (see Table 3 of Cohen 1979). Therefore, we are able to 
classify the spectrum of CGCS 5926 as C(6,2). We note that the use of the 
diagnostic $D$ of Cohen et al. (1996) instead of the temperature index $T$ 
for the determination of the parameter $m$ gives us a somewhat lower value 
(around 5.5), implying a slightly higher temperature and thus an earlier 
spectral type.

The results of our photometry of CGCS 5926 are reported in Table 1. These 
indicate that the source shows a variability of amplitude $\Delta m \sim$ 
0.3 mag on timescales of tens of days. Moreover, the $V-R_{\rm C}$ and 
$R_{\rm C}-I_{\rm C}$ color indices of CGCS 5926 seem to get smaller as 
the $V$-band magnitude decreases; that is, the star gets bluer with 
increasing brightness. Therefore we carried out a periodicity search on 
the photometric data using the Fourier code of Deeming (1975). A single, 
strong probability peak was found at a period of 151 days. The ephemeris 
we obtained with this period, expressed in Heliocentric Julian Days 
(HJDs), provides the following times of maxima in the $V$ band:

\begin{equation}
\mathrm{T}_{\mathrm{max},V} ~ = ~ 2455164 (\pm 3) ~ + ~ (151 \pm 2) \times E~,
\end{equation}

\noindent
where $E$ is an integer number; the errors are at 3$\sigma$ confidence 
level. The $V$-band and the $V-I_{\rm C}$ color light curves of CGCS 5926 
folded onto this ephemeris are presented in Fig.~4. Their shape, and 
the fact that the star gets bluer at maximum and redder at minimum optical 
brightness, suggest that this variability traces the radial pulsation of 
this carbon star (see e.g. Wallerstein \& Knapp 1998). 

Because of the relatively short time baseline covered by our observations 
($\sim$400 days, that is, a bit more than two pulsation cycles), we cannot 
build a mean light curve (i.e. one averaged over several and well covered 
pulsation cycles) for the star. Thus, the only feasible approach to 
estimate the above errors on the period and the epoch was to impose that 
the lightcurve is the simplest and smoothest possible, with a shape similar 
to that of other objects of that variability class (in this case, a rise 
to maximum faster than the decline to minimum). Besides, we stress that 
these stars usually display a significant variability of the period and 
epoch with time around mean values (possibly due to beating of multiple 
periodicities; see Wallerstein \& Knapp 1998 and references therein).

\begin{table*}
\caption[]{Log and results of the optical $BVR_{\rm C}I_{\rm C}$ 
photometry of star CGCS 5926 presented in this paper (see text for 
details).}
\begin{center}
\begin{tabular}{cccccc}
\noalign{\smallskip}
\hline
\hline
\noalign{\smallskip}
   HJD-2450000 & $V$ & $B$$-$$V$ & $V$$-$$R_{\rm C}$ & $R_{\rm C}$$-$$I_{\rm C}$ & $V$$-$$I_{\rm C}$ \\
\noalign{\smallskip}
\hline
\noalign{\smallskip}
 5119.6871 & 14.949$\pm$0.007 & 2.58$\pm$0.04 & 1.789$\pm$0.008 & 1.639$\pm$0.004 & 3.432$\pm$0.007 \\  
 5119.6978 & 14.978$\pm$0.007 & 2.73$\pm$0.05 & 1.819$\pm$0.008 & 1.631$\pm$0.004 & 3.453$\pm$0.007 \\ 
 5120.6184 & 14.948$\pm$0.008 & 2.72$\pm$0.06 & 1.852$\pm$0.009 & 1.626$\pm$0.004 & 3.479$\pm$0.008 \\ 
 5120.6294 & 14.935$\pm$0.007 & 2.75$\pm$0.05 & 1.824$\pm$0.008 & 1.629$\pm$0.005 & 3.458$\pm$0.008 \\ 
 5131.8875 & 14.881$\pm$0.008 & 2.68$\pm$0.05 & 1.794$\pm$0.009 & 1.635$\pm$0.006 & 3.432$\pm$0.010 \\ 
 5131.8974 & 14.916$\pm$0.008 & 2.76$\pm$0.06 & 1.827$\pm$0.009 & 1.643$\pm$0.006 & 3.475$\pm$0.009 \\ 
 5136.7797 & 14.881$\pm$0.012 & 2.95$\pm$0.15 & 1.800$\pm$0.013 & 1.622$\pm$0.007 & 3.427$\pm$0.014 \\ 
 5136.7897 & 14.869$\pm$0.012 & 2.70$\pm$0.09 & 1.798$\pm$0.013 & 1.626$\pm$0.007 & 3.424$\pm$0.014 \\ 
 5138.8442 & 14.863$\pm$0.013 & 2.82$\pm$0.11 & 1.794$\pm$0.014 & 1.627$\pm$0.007 & 3.424$\pm$0.014 \\ 
 5138.8542 & 14.876$\pm$0.013 & 2.78$\pm$0.10 & 1.835$\pm$0.014 & 1.617$\pm$0.007 & 3.453$\pm$0.014 \\ 
 5161.8343 & 14.688$\pm$0.009 & 2.87$\pm$0.06 & 1.745$\pm$0.012 & 1.586$\pm$0.009 & 3.328$\pm$0.011 \\ 
 5161.8542 & 14.665$\pm$0.012 & 3.06$\pm$0.09 & 1.757$\pm$0.014 & 1.555$\pm$0.009 & 3.305$\pm$0.014 \\ 
 5167.8041 & 14.693$\pm$0.016 & 2.80$\pm$0.20 & 1.751$\pm$0.016 & 1.567$\pm$0.010 & 3.312$\pm$0.018 \\ 
 5167.8223 & 14.684$\pm$0.015 & 2.82$\pm$0.17 & 1.750$\pm$0.016 & 1.571$\pm$0.009 & 3.318$\pm$0.016 \\ 
 5192.7497 & 14.844$\pm$0.015 & 2.71$\pm$0.11 & 1.750$\pm$0.013 & 1.636$\pm$0.010 & 3.389$\pm$0.016 \\ 
 5444.8529 & 14.824$\pm$0.006 & 2.67$\pm$0.04 & 1.798$\pm$0.007 & 1.587$\pm$0.006 & 3.370$\pm$0.007 \\
 5444.8675 & 14.814$\pm$0.006 & 2.71$\pm$0.04 & 1.775$\pm$0.007 & 1.588$\pm$0.005 & 3.353$\pm$0.008 \\
 5445.8451 & 14.819$\pm$0.006 & 2.76$\pm$0.05 & 1.774$\pm$0.007 & 1.608$\pm$0.006 & 3.372$\pm$0.008 \\
 5445.8597 & 14.824$\pm$0.006 & 2.66$\pm$0.05 & 1.781$\pm$0.007 & 1.608$\pm$0.005 & 3.379$\pm$0.007 \\
 5477.0016 & 14.752$\pm$0.010 & 2.84$\pm$0.07 & 1.772$\pm$0.009 & 1.590$\pm$0.006 & 3.353$\pm$0.010 \\
 5502.7452 & 14.809$\pm$0.008 & 2.98$\pm$0.04 & 1.770$\pm$0.008 & 1.624$\pm$0.011 & 3.398$\pm$0.014 \\
 5505.7462 & 14.820$\pm$0.008 & 2.96$\pm$0.05 & 1.785$\pm$0.010 & 1.631$\pm$0.007 & 3.421$\pm$0.009 \\
 5505.7650 & 14.812$\pm$0.010 & 2.92$\pm$0.05 & 1.778$\pm$0.012 & 1.642$\pm$0.009 & 3.403$\pm$0.014 \\
 5512.7498 & 14.841$\pm$0.007 & 2.81$\pm$0.04 & 1.782$\pm$0.009 & 1.619$\pm$0.007 & 3.407$\pm$0.009 \\
 5513.7223 & 14.833$\pm$0.007 & 2.74$\pm$0.05 & 1.775$\pm$0.010 & 1.612$\pm$0.008 & 3.393$\pm$0.009 \\
 5513.7413 & 14.833$\pm$0.009 & 2.86$\pm$0.05 & 1.780$\pm$0.011 & 1.625$\pm$0.008 & 3.413$\pm$0.012 \\
 5514.7398 & 14.838$\pm$0.007 & 2.77$\pm$0.05 & 1.783$\pm$0.010 & 1.615$\pm$0.007 & 3.403$\pm$0.009 \\
 5516.6971 & 14.810$\pm$0.008 & 2.67$\pm$0.06 & 1.753$\pm$0.010 & 1.606$\pm$0.007 & 3.366$\pm$0.012 \\
 5516.7161 & 14.820$\pm$0.008 & 2.84$\pm$0.08 & 1.760$\pm$0.010 & 1.617$\pm$0.009 & 3.386$\pm$0.010 \\
 5517.7169 & 14.812$\pm$0.009 & 2.73$\pm$0.07 & 1.769$\pm$0.009 & 1.614$\pm$0.009 & 3.395$\pm$0.010 \\
 5517.7369 & 14.842$\pm$0.008 & 2.80$\pm$0.08 & 1.792$\pm$0.011 & 1.601$\pm$0.008 & 3.395$\pm$0.011 \\
\noalign{\smallskip}
\hline
\noalign{\smallskip}
\end{tabular}
\end{center}
\end{table*}

\begin{figure}
\hspace{-.8cm}
\psfig{figure=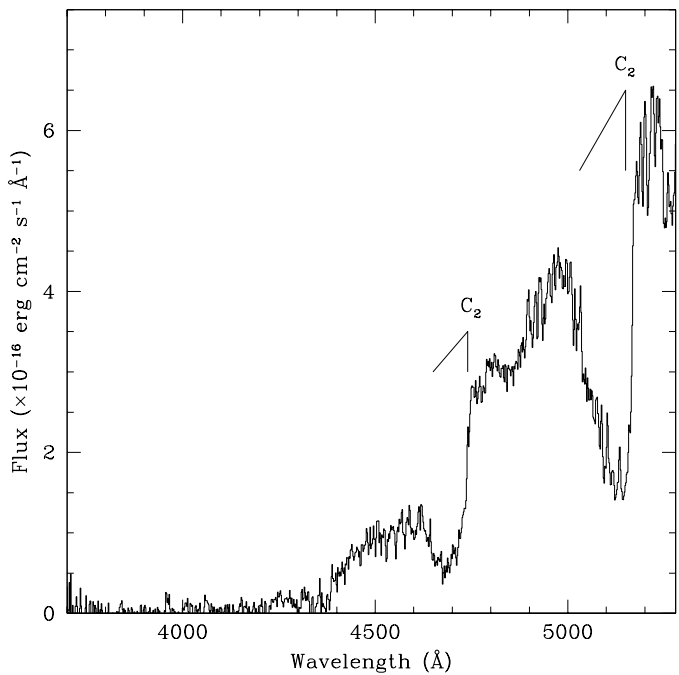,width=10.5cm}
\vspace{-1.2cm}
\caption[]{3700--5300 \AA~optical spectrum of star CGCS 5926 obtained 
with the Cassini telescope on 9 August 2011. The bluest C$_2$ Swan bands are 
labeled. No signal is detected from the object blueward of $\sim$4200 \AA.}
\end{figure}

\begin{figure}
\psfig{figure=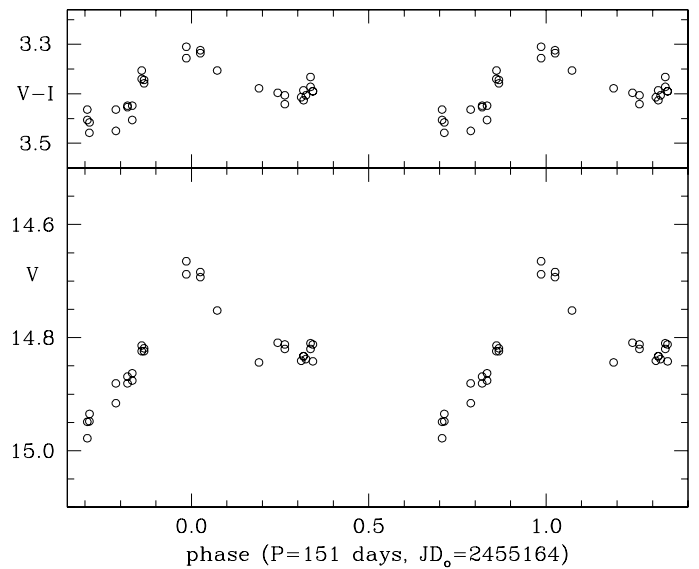,width=9cm,angle=0}
\caption[]{$V$ {\it (lower panel)} and $V-I_{\rm C}$ {\it (upper panel)} 
light curves of CGCS 5926 folded with a period of 151 days and with zero 
phase at Julian Day 2455164. We identify this as the radial pulsation 
periodicity of the star, which gets its bluest color at maximum 
brightness.}
\end{figure}

One can then estimate the distance to CGCS 5926 in the following manner. 
Assuming an average magnitude $V \sim$ 14.8 for the source, its $JHK_{\rm 
s}$ magnitudes as reported in the 2MASS catalog (see Section 1) and the 
intrinsic color indices for carbon stars as reported in Table 6 of Ducati 
et al. (2001), one gets an average optical reddening $A_V \sim$ 3.8 mag, 
which corresponds to a color excess $E(B-V) \sim$ 1.23 mag considering 
the Galactic absorption law of Cardelli et al. (1989) and a 
total-to-selective extinction ratio $R_V$ = 3.1 (Rieke \& Lebofski 1985). 
This implies that the intrinsic magnitude of the star is $V_0 \sim$ 11.0.

Therefore, given that a carbon star of spectral type C(6,2) has an 
absolute magnitude M$_V \approx $ $-$2.6 (Cohen 1979), we obtain a 
distance $d \sim$ 5.2 kpc. This would locate CGCS 5926 in the far side of 
the outer part of the Perseus Arm of the Galaxy (see e.g. Fig. 1 of Leitch 
\& Vasisht 1998). The Galactic line-of-sight 
reddening in the direction of CGCS 5926 is $E(B-V)$ = 1.33 mag according 
to the maps of Schlegel et al. (1998). This value compares well with our 
reddening estimate for the source, thus indicating that this star is 
behind that Galactic Arm and that the observed reddening is likely due to 
interstellar absorption only, with no substantial contribution from 
material local to the source. We nevertheless note that the tabulated 
Galactic absorption maps should be treated with some degree of caution for 
objects which have the line of sight along the Galactic Plane, such as the 
present case (for which the Galactic latitude is $b$ = +0$\fdg$96).

We remark that the above approach rules out the possibility that CGCS 5926 
is a red supergiant of luminosity class I: indeed, in this case, its 
absolute magnitude would be M$_V \sim $ $-$5.6 (Lang 1992), which would 
place the object at an uncomfortably large distance of $\sim$20 kpc, thus
well outside the Galaxy given the Galactic longitude of the source ($l$ 
= 115$\fdg$5).

No X--ray source was detected in the XRT pointing, either at the optical 
position of CGCS 5926 or within the {\it ROSAT} error circle. Using the 
bayesian approach of Kraft et al. (1991), we get a 3-$\sigma$ confidence 
level upper limit count rate of 2$\times$10$^{-3}$ s$^{-1}$ in the 0.3--10 
keV energy band. This, assuming a Crab-like spectrum, corresponds to an 
observed flux limit of $\sim$9$\times$10$^{-14}$ erg cm$^{-2}$ s$^{-1}$ in 
this energy range. Likewise, no UV source was detected in coincidence of 
the target down to a 3-$\sigma$ limit of 21.8 mag in the $UVM2$ band.

The above {\it Swift}/XRT upper limit in X--rays can therefore be compared 
to the detection in the {\it ROSAT} faint source catalog: here the {\it 
ROSAT} source is reported with a 0.1--2.4 keV count rate of 
(1.96$\pm$0.76)$\times$10$^{-2}$ s$^{-1}$. This, assuming again a 
Crab-like spectrum, implies a flux of $\sim$1$\times$10$^{-13}$ erg 
cm$^{-2}$ s$^{-1}$. It can be seen that the {\it ROSAT} spectral coverage 
is basically a subset of the XRT one; moreover, the contribution of the 
X--ray emission below 0.3 keV to the total flux should not be relevant 
given the non-negligible line-of-sight absorption toward CGCS 5926 
apparent from the optical data, which converts into a hydrogen column 
density $N_{\rm H} \sim$ 6.8$\times$10$^{21}$ cm$^{-2}$ if one uses the 
empirical formula of Predehl \& Schmitt (1995). The above figures thus 
suggest that, if the {\it ROSAT} detection is real, the source should 
possess variability at X--rays as well. We will discuss this in detail in 
the next section.

These fluxes, using the above distance estimate for CGCS 5926, thus imply 
an X--ray luminosity of $\approx$3$\times$10$^{32}$ erg s$^{-1}$ for the 
{\it ROSAT} measurement and an upper limit of comparable magnitude in the 
case of the {\it Swift}/XRT pointing.

To conclude this section, we note that the area of the sky containing CGCS 
5926 was observed twice within the {\it XMM-Newton} Slew 
Survey\footnote{see also {\tt 
http://xmm.esac.esa.int/external/xmm\_products/ \\
slew\_survey/upper\_limit/uls.shtml}} 
(Saxton et al. 2008) on 6 January 2007 and on 6 February 2009 for 7 and 9 
seconds, respectively. For both observations, only a loose 3-$\sigma$ 
upper limit of $\approx$2.5$\times$10$^{-12}$ erg cm$^{-2}$ s$^{-1}$ in 
the 0.2--12 keV band was obtained.

\section{Discussion}

Our optical spectroscopy of star CGCS 5926 confirms that this object can 
indeed be classified as a red giant carbon star; the optical photometry 
presented in this paper indicates that the source is slightly and 
periodically variable at least on timescales of a few months, which 
is also typical of this class of red giants and likely due to 
stellar pulsations (e.g. Wallerstein \& Knapp 1998). Our combination of 
optical and NIR photometry allowed us to determine a distance to the 
source $d \sim$ 5.2 kpc.

As shown in the previous section, no substantial continuum emission 
was detected in the blue part of the optical spectrum of CGCS 5926 
(Fig.~3). The absence of a continuum excess of the order of few tenths of 
magnitude in the blue and in the ultraviolet, which is sometimes observed 
in SyXBs (see e.g. Gaudenzi \& Polcaro 1999), does not in itself rule out 
the classification of CGCS 5926 as a new member of this class of objects 
because of the large interstellar absorption we determine along the line 
of sight to this star. Indeed, our estimate implies about 6 magnitudes of 
attenuation in the $U$ band assuming the law of Cardelli et al. (1989).

Similarly, the source was not detected in X--rays by {\it Swift}/XRT down 
to a limit of 3$\times$10$^{32}$ erg s$^{-1}$ in the 0.3--10 keV band, 
which again may cast doubts on the possibility that this source can be a 
SyXB. Actually, the lowest fluxes detected in X--rays from this type of 
source (Masetti et al. 2002, 2007; Nespoli et al. 2010), are of the order 
of the above limit for CGCS 5926. Thus the X--ray nondetection does not in 
itself allow us to rule out that this star is the optical counterpart of a 
SyXB.

A possible explanation for the lack of an X--ray detection from the source 
is that we observed it during an eclipse of the compact accretor.
Indeed a carbon star, being at the tip of the AGB, has a very large radius 
(e.g. Cohen 1979). Thus, when seen from the companion, it fills a large 
fraction of its sky, and as a matter of fact one third of all well studied 
symbiotic stars are eclipsing systems (Miko\l ajewska 2003). The geometric 
transit of the compact companion behind the cool giant can therefore take 
up to months, and the X--ray nondetection with {\it Swift} could be 
justified by the accreting component undergoing an eclipse at that time.

To explore this possibility we can first evaluate the size of the orbit in 
the case of accretion via the inner Lagrangian point. In this occurrence, 
it is generally assumed that the size of the Roche lobe of a star in a 
binary system is characterized by its radius $R_{\rm L}$, defined as the 
radius of a sphere with the same volume of the Roche lobe. This quantity 
is related to the orbital semimajor axis $a$ according to the 
approximation of Eggleton (1983):

\begin{equation}
\frac{R_L}{a} = \frac{0.49~ q^{2/3} }{0.6~ q^{2/3}+~ \mathrm{ln}~(1 + q^{1/3})}~,
\end{equation}

\noindent
where $q$ is the mass ratio of the system. Assuming that the mass of the 
giant companion $M_c$ is $\approx$1.4 $M_\odot$ (we here use an average 
value; see Wallerstein \& Knapp 1998) and that the accretor is a NS (thus 
with a mass $M_{\rm NS} \approx$ 1.4 $M_\odot$), we obtain that $q \approx$ 
1. Moreover, equating $R_{\rm L}$ to the typical radius of a C(6,2) giant 
carbon star ($R \approx$ 470 $R_\odot$: Cohen 1979), we get from Eq. (2) 
that $a \approx$ 1200 $R_\odot$. This, by using Kepler's third law, implies 
an orbital period of $\approx$3000 days and an inclination $i \ga$ 
70$^\circ$ for the occurrence of an X--ray eclipse. This estimate, by the 
way, reinforces the fact that the periodicity we determine in the optical 
light curve is not of orbital nature but rather has its origin in the 
pulsation of the stellar structure of CGCS 5926.

However, in the present case the matter transfer onto the compact object 
is unlikely to be occurring via Roche lobe overflow as this would produce 
X--ray emission which is $\sim$5 orders of magnitude more intense than the 
upper limit that we obtained with {\it Swift}/XRT (as for example in GX 
1+4: Chakrabarty \& Roche 1997); our very detection of regular radial 
pulsations reinforces this assumption. Therefore, the above figures 
should rather be considered as hard lower limits for orbital period and 
inclination.

The mass transfer would thus take place via a stellar wind, which makes 
the overall accretion mechanism much less efficient and the X--ray 
luminosity much lower (but see Mohamed \& Podsiadlowski 2007, 2011): 
in this case the orbit will be much wider.

Assuming thus the orbital period to be larger by at least a factor of 4 
with respect to the above estimate, and again using Kepler's third law, we 
get $a \ga$ 3100 $R_\odot$: this means that, in order to observe an eclipse 
from such a system, its inclination should be $\ga$80$^\circ$. In this case, 
the eclipse would last not more than $\sim$75 days, which is a tiny fraction 
(less than 1\%) of the orbital period. This however tells us that, even if 
with admittedly a low level of probability, the hypothesis of an eclipsing 
SyXB cannot be completely discarded.

A less contrived scenario is possibly the one in which we consider that 
the source is highly variable in X--rays, as seen in other SyXBs. For 
instance, the X--ray fluxes of 4U 1700+24 (Masetti et al. 2002) and 4U 
1954+31 (Masetti et al. 2006b) span 2 orders of magnitude; likewise, and 
in a more extreme manner, that of IGR J16358$-$4726 spans a dynamical 
range of 10$^4$ (Patel et al. 2007; Nespoli et al. 2010). This can be 
qualitatively explained by accretion from an inhomogeneous stellar wind 
(possibly modulated by stellar pulsations) coming from a red giant star, 
possibly coupled with an elliptical orbit of the accretor.

If we assume that the X--ray luminosity coming from CGCS 5926 and detected 
with {\it ROSAT} is due to accretion from a red giant stellar wind, an 
accretion rate onto the compact star of $\dot{M} \approx$ 
3$\times$10$^{-14}$ $M_\odot$ yr$^{-1}$ is inferred. This can be 
interpreted by assuming a typical mass loss rate via stellar wind of a 
giant carbon star ($\dot{M} \approx$ 10$^{-7}$ $M_\odot$ yr$^{-1}$, and 
sometimes higher; e.g. Wallerstein \& Knapp 1998) and an accretion 
efficiency $\eta$ = 10$^{-4}$ which is quite normal for a compact object 
accreting from a stellar wind (e.g. Frank et al. 1992). In the case that 
the accreting object is a NS, the remaining difference between the above 
two values of $\dot{M}$ can possibly be accounted for by a partial 
inhibition of the accretion due to the ``propeller effect" (Illarionov \& 
Sunyaev 1975), according to which the magnetosphere of the NS acts as a 
barrier to accretion of matter onto the NS surface. It is however noted 
that this star is not detected in the IRAS far-infrared (FIR) source 
catalogue (IRAS 1986). Given the direct correlation between the mass loss 
in carbon stars and their FIR emission (e.g. Wallerstein \& Knapp 1998), 
we may expect that the $\dot{M}$ of CGCS 5926 is actually substantially 
lower than 10$^{-7}$ $M_\odot$ yr$^{-1}$. This would make the propeller 
effect hypothesis more viable to explain the low level of X--ray emission 
from CGCS 5296 assuming that this star is part of a SyXB system.

It is prudent to verify whether the X--ray emission detected with 
{\it ROSAT} is indeed associated with CGCS 5926 rather than, for 
instance, other objects within the X--ray error circle (see Fig. 1). In 
order to check this, we here evaluate the probability of finding a 
late-type giant star in a 20$''$-radius circle at these Galactic 
latitudes. To this aim, we first made extensive catalogue searches using 
the SIMBAD and VIZIER databases (see for instance Maehara \& Soyano 1999 
and references therein). We found that the total number of known Carbon 
stars in our Galaxy is less than 2000 (including suspected and tentative 
ones). Even assuming the extreme case in which they are all concentrated 
in a strip on the Milky Way plane with Galactic latitude between $b$ = 
$-$5$^\circ$ and $b$ = +5$^\circ$, we get a density of $\la$0.5 carbon 
stars per square degree; this in turn implies a probability less than 
5$\times$10$^{-5}$ to have one such star in a circle of radius 20$''$.

A more conservative estimate can be obtained using the `Besan\c{c}on' 
Galactic model (Robin et al. 2003). This population synthesis description 
of the Galaxy returned a total probability of 5.0$\times$10$^{-3}$ that, 
within a circle of radius 20$''$ around the position of the {\it ROSAT} 
source 1RXS J234545.9+625256, we find an M-type red giant of luminosity 
class I, II or III. Thus, the probability of finding a red giant star 
within the {\it ROSAT} error circle at the Galactic coordinates of CGCS 
5926 is a number between the two estimates above, and thus of the order of 
$\approx$10$^{-4}$. This suggests that the chance probability of the 
positional coincidence between the star CGCS 5926 and the {\it ROSAT} 
source 1RXS J234545.9+625256 is quite low, although not vanishing. We 
thus remain with an admittedly very small possibility that the two objects 
are actually not correlated.

One may thus wonder whether other classes of Galactic or extragalactic X--ray 
sources may display this variable high-energy behaviour, so that one of the 
other fainter optical objects within the {\it ROSAT} error circle in Fig. 1 
may be the actual counterpart of the X--ray emitter. For instance, the 
{\it ROSAT} detection may have been produced because of magnetic activity 
from Galactic objects such as a `flare star' (e.g. Pettersen 1989), or 
by the outburst of a very faint X--ray transient (VFXT) LMXB (see e.g. 
Degenaar \& Wijnands 2009). 

The latter objects are however mostly associated with old (bulge) stellar 
population, so we do not expect to find one of them far from the inner parts 
of the Galaxy. Moreover, and most importantly, these transients reach X--ray 
luminosities of at least 10$^{34}$ erg s$^{-1}$ (e.g., Degenaar \& Wijnands 
2009), which means that the {\it ROSAT} source would lie at a distance of at 
least 29 kpc, that is, well outside the Galaxy. Therefore, the VFXT 
hypothesis is not tenable for the {\it ROSAT} detection.

The possibility that one of the other sources within the {\it ROSAT} error 
circle (see Fig. 1) is a dwarf, late-type flare star of UV Cet type cannot
of course be excluded a priori, especially given that we have no specific 
information on any of them. However, using the USNO-A2.0\footnote{Available at \\
{\tt http://archive.eso.org/skycat/servers/usnoa/}} magnitudes of
these sources we find that the brighter ones are too blue to be M-type dwarfs,
while the fainter ones lie at distances $d \ga$ 700 pc (assuming an absolute 
magnitude M$_R \approx$ +10; e.g., Lang 1992); this would mean that the 
{\it ROSAT} detection implies an X--ray luminosity $\ga$3$\times$10$^{30}$ 
erg s$^{-1}$, which is extreme for an eruption of a flare star (see for 
instance Pandey \& Singh 2008). We caution the reader that the USNO-A2.0 
data can have systematic uncertainties of a few tenths of magnitude (see 
Masetti et al. 2003), and that in the above considerations we did not take 
into account the interstellar reddening (which would move the above upper 
limit on the distance closer to earth and would relax the X--ray luminosity 
constraint). However, we do not think that the latter is a real issue because 
we expect that the bulk of the absorption lies within the Perseus arm, that 
is, at a distance larger than $\sim$2 kpc from earth according to Fig. 1 of 
Leitch \& Vasisht (1998). Thus, although more reasonable than the VFXT one, 
we deem the flare star interpretation for the {\it ROSAT} detection still
unlikely.

We can in any case definitely exclude a flare star nature for CGCS 5926
given that it is an evolved star, whereas those are typically young
and fast rotating stellar objects; moreover, these objects have 
X--ray luminosities which rarely exceed 10$^{30}$ erg s$^{-1}$ (e.g.,
Pandey \& Singh 2008), which means 2 orders of magnitude less that the 
{\it ROSAT} detection in case it was emitted by CGCS 5926. Similarly, this
red star cannot be associated with a very faint X--ray transient as these 
systems are in general `normal' LMXBs with a late-type (or sometimes a
degenerate) dwarf as donor star.

As regards the possibility of an extragalactic object being responsible 
for the X--ray emission detected with {\it ROSAT}, we consider that the most 
likely case would be that of a background active galactic nucleus (AGN). 
In order to evaluate the probability of the occurrence of such a source 
within the {\it ROSAT} error circle we use the
AGN density relation of Cappelluti et al. (2007). We see that the chance of
finding a field object with a {\it ROSAT} X--ray flux larger than 
$\approx$10$^{-13}$ erg cm$^{-2}$ s$^{-1}$ in a 20$''$-radius circle is 
$\approx$2$\times$10$^{-4}$: this again implies a very low, although not
completely negligible, chance coincidence probability that the {\it ROSAT}
detection was due to an AGN located beyond the Galactic Plane.

Finally, the only alternative explanation that may come to mind is 
that the low confidence level detection with {\it ROSAT} is actually 
spurious, and that CGCS 5926 does not emit appreciable X--rays, especially 
above 2 keV. This may suggest that no accretion onto a compact companion 
is taking place. If this interpretation is correct, CGCS 5926 can be 
discarded as a further possible SyXB. Moreover, as our data do not give us 
any hint about this, again there may possibly be no companion at all for 
CGCS 5926, making it a single, isolated carbon star. In this case, we 
remark that our X--ray flux limit would still be compatible with the 
coronal emission from a late-type giant star, this being generally about 3 
orders of magnitude lower than our result (H\"unsch et al. 1998).

\section{Conclusions}

We performed a multiwavelength study of star CGCS 5926 in X--ray, 
ultraviolet, optical and NIR bands using new observations as well as 
archival data when available. This was done in order to investigate the 
possible SyXB nature of this source due to its positional proximity with 
the {\it ROSAT} X--ray source 1RXS J234545.9+625256.

The optical data confirm that the object is a carbon star likely located 
at $\sim$5.2 kpc from earth and show the presence of radial pulsations 
with a periodicity of 151 days. 
No detection of X--ray or ultraviolet emission was obtained from the star 
in 2010, nor was any excess detected in the blue part of its optical 
spectrum. While all of this casts doubts on the former {\it ROSAT} detection 
and even on the possibility that this source is a SyXB (or even a binary 
system at all), our multiwavelength data analysis does not rule out either 
of these possibilities. Alternative interpretations for the variable
X--ray emission, such as an interloping flare star, a field VFXT or a 
background AGN were found to be highly unlikely or were definitely ruled out. 
Thus, we remain with the two possibilities above: either CGCS 5926 is 
indeed a SyXB, or the {\it ROSAT} source is spurious.

Future spectrophotometric optical and NIR monitoring over a long time 
baseline (years) will help determine if indeed CGCS 5926 is part of a 
binary system, while deeper, longer, and higher sensitivity X--ray 
observations with satellites such as {\it XMM-Newton} or {\it Chandra} 
will provide a possible detection of, or at least tighter upper limits on, 
the high-energy emission from this source.

\begin{acknowledgements}

We are grateful to the following people for their assistance in the
acquisition of the optical spectra of CGCS 5926: Alessandro Siviero
(Asiago), Gianni Cetrulo and Antonio Englaro (La Polse di Cougnes), 
Alberto Milani (Campo de' Fiori), and Silvia Galleti and Roberto Gualandi 
(Loiano). We also thank Mauro Dadina and Isabella Pagano for discussions, 
and the anonymous referee for useful remarks which helped us to improve 
the quality of this paper. 
This research has made use of the ASI Science Data Center Multimission 
Archive; it also used the NASA Astrophysics Data System Abstract Service 
and the NASA/IPAC Infrared Science Archive, which are operated by the Jet 
Propulsion Laboratory, California Institute of Technology, under contract 
with the National Aeronautics and Space Administration. 
This publication made use of data products from the Two Micron All 
Sky Survey (2MASS), which is a joint project of the University of 
Massachusetts and the Infrared Processing and Analysis Center/California 
Institute of Technology, funded by the National Aeronautics and Space 
Administration and the National Science Foundation.
This research has also made extensive use of the SIMBAD and VIZIER 
databases operated at CDS, Strasbourg, France. 
NM acknowledges financial contribution from the ASI-INAF agreement No. 
I/009/10/0. 
KLP and JPO acknowledge the support of the UK Space Agency.
SS acknowledges partial support from NSF and NASA grants to ASU.
\end{acknowledgements}

\end{document}